\documentclass[twocolumn,twocolappendix]{aastex631}
\usepackage{amssymb}	
\usepackage{hyperref}

\usepackage{graphicx}	
\usepackage{amsmath}	
\DeclareMathOperator{\sech}{sech}
\usepackage{amssymb}	

\shorttitle{Halo Spin Effect on Bar X-shapes}

\begin{document}
\title[Halo Spin Effect on Bar X-shapes]{Effects of Inner Halo Angular Momentum on the Peanut/X-shapes of Bars}

\correspondingauthor{Sandeep Kumar Kataria, Juntai Shen}
\email{skkataria.iit@gmail.com, jtshen@sjtu.edu.cn}

\author[0000-0003-3657-0200]{Sandeep Kumar Kataria}
\affiliation{Department of Astronomy, School of Physics and Astronomy, Shanghai Jiao Tong University, 800 Dongchuan Road, Shanghai 200240, China}
\affiliation{Key Laboratory for Particle Astrophysics and Cosmology (MOE) / Shanghai Key Laboratory for Particle Physics and Cosmology, Shanghai 200240, China
}

\author[ 0000-0001-5604-1643]{Juntai Shen}
\affiliation{Department of Astronomy, School of Physics and Astronomy, Shanghai Jiao Tong University, 800 Dongchuan Road, Shanghai 200240, China}
\affiliation{Key Laboratory for Particle Astrophysics and Cosmology (MOE) / Shanghai Key Laboratory for Particle Physics and Cosmology, Shanghai 200240, China
}

\begin{abstract}

Cosmological simulations show that dark matter halos surrounding baryonic disks have a wide range of angular momenta, which is measured by the spin parameter ($\lambda$). In this study, we bring out the importance of inner angular momentum($<$30 kpc), measured in terms of the halo spin parameter, on the secular evolution of bar using N-body simulations. We have varied the halo spin parameter $\lambda$ from 0 to 0.1, for co-rotating (prograde) spinning halos and one counter-rotating (retrograde) halo spin ($\lambda$=-0.1) with respect to disk. We report that as the halo spin increases the buckling is also triggered earlier and is followed by a second buckling phase in high spin halo  models. The timescale for the second buckling is significantly longer than the first buckling. We find that bar strength does not reduce significantly after the buckling in all of our models which provide new insights about the role of inner halo angular momentum unlike previous studies. Also, the buckled bar can still transfer significant angular momentum to the halo in  the secular evolution phase, but it reduces with increasing halo spin. In the secular evolution phase, the bar strength increases and saturates to nearly equal values for all the models irrespective of halo spin and the sense of rotation with respect to disk. The final boxy/peanut shape is more pronounced ($\sim$20 $\%$) in high spin halos having higher angular momentum in the inner region compare to non-rotating halos. We explain our results with angular momentum exchanges between disk and halo.

\end{abstract}

\keywords{dark matter – galaxies: spiral – galaxies: evolution – galaxies: kinematics and dynamics – methods: numerical}


\section{Introduction} \label{sec:intro}
Cosmological simulations show that dark matter halos play an important role in galaxy formation and evolution \citep{Springel.etal2005_Millenium,Schaye.et.al.2015}. Dark matter halos grow gradually from the initial primordial perturbations, given the long range nature of gravity, and result in the formation of huge potential wells in which baryonic matter condenses \citep{White.Rees.1978}. As a result dark matter halos in cosmological simulations come with various properties such as different shapes, sizes, angular momenta and concentration. Furthermore, using isolated disk galaxy evolution studies it has been shown that the various dark matter properties like concentration and shape affect the morphological evolution of the stellar disk via global instabilities \citep{Athanassoula.Misiriotis.2002,Athanassoula.2003,Athanassoula.et.al.2013,Kumar.et.al.2022}.

    The origin of halo angular momentum has been shown to be the result of tidal interactions between  neighbouring galaxies as well as neighbouring filament structures which causes perturbations in the galaxy potentials; this is famously known as Tidal Torque Theory \citep{Hoyle.1949,Peebles.1969,Sciama.1955,Doroshkevich.1970,White.1984,Barnes.Efstathiou.1987,schafer.2009}. Recently it has been shown that merger histories of a given halo can significantly affect it's angular momentum \citep{Maller.et.al.2002,Vitvitska.et.al.2002} which is slightly different from what is predicted by Tidal Torque Theory. The rotation of these halos, which is the proxy of it's angular momentum is measured by a parameter called the halo spin. It has been shown that the lognormal distribution of the halo spin peaks around 0.035 \citep{Bullocketal.2001, Hetznecker.Burkert.2006,Bett.et.al.2007,Knebe.Power.2008,Ishiyama.et.al.2013,Zjupa.Springel.2017} where spin is defined as:

    \begin{equation}
         \lambda=\dfrac{J}{\sqrt{2GMR}}  
         \label{eq:lambda}
    \end{equation}
       
    Here J is the magnitude of the specific angular momentum of the halo, M is the mass of the halo within the virial radius and R is the virial radius of the halo. It is well known that disks with global instabilities like bars can transport angular momentum to the surrounding halo during its evolution in isolation \citep{Debattista.Sellwood.2000,Athanassoula.2003,Martinez-valpuesta.2006,villa-varga.et.al.2010,Sellwood.2016,Kataria.Das.2018,Lokas.2019}. Therefore the initial angular momentum in the dark matter halo, which is a proxy for halo spin, affects the process of angular momentum transfer from disk to halo. 
    
    A recent study by \cite{Kanak.Saha.Naab.2013} show that dark matter halos with higher spin trigger the onset of bar formation earlier than low spinning halos which resonates with previous theoretical studies \citep{Weinberg.1985}. Furthermore, \cite{Longetal.2014} show that models with increasing halo spin tend to weaken the bar in the secular evolution phase which is followed by strong buckling of the bar. These studies clearly show that angular momentum transfer between the halo and disk is significantly affected by varying halo spins, which in turn affects the formation and evolution of bars in the disks.  The effect of halo shapes on increasing halo spin has also been studied \citep{Collieretal.2018}. Along with this, the effect of halos counter-rotating with respect to the disk has been explored by \cite{Collieretal.2019},  and they show that there is a delay in bar formation with the increasing counter-rotating spin of halos. All these studies indicate that there is rich dynamics between spinning halos and disks which certainly depend on the initial angular momentum of the dark matter halo. 
    
    In this study we focus on the effect of the inner halo angular momentum which has a close  proximity with the disk, on bar formation and evolution. We measure the angular momentum difference of all of our galaxy models in terms of halo spin, assuming that the spin is correlated with increasing inner halo angular momentum. We have varied the inner halo angular momentum by varying prograde/retrograde orbits in the inner halo. Given the role of halo spin on the evolution of  bars, we will also focus on the buckling of bars. It is well known that bars undergo buckling instability during their evolution and just after they gain their peak strength \citep{Combes.et.al.1990,Pfenniger.Friedli.1991,Raha.etal.1991,Martinez-valpuesta.2006,Shen.et.al.2010,Collier.2020}, which leads to decrease in bar strength. Previous studies \citep{Longetal.2014,Collieretal.2018} show that strong buckling along with increase in halo spin leads to weakening of bars in the secular evolution phase.
   The next interesting question to answer will be ``How does weak and intermediate buckling, which does not affect bar strength significantly, respond to different values of halo spin?"  So in this study we address the following questions ~:~1) How is buckling strength affected by increasing halo spin? 2) How does the peanut/X-shape of a bar depend on the halo spin ? 3) How is the thickening of bars affected by different mechanisms \citep{Sellwood.Gerhard.2020} ?

    In this article we  construct galaxy models with increasing halo spin ($\lambda \approx 0-0.1$) which are co-rotating with respect to the disk, as well as one counter-rotating halo model. In these models, the spin of halos is controlled by varying inner halo angular momentum only. We evolve these models to see the effect of halo spin on the  evolution of bars, like buckling and the formation of boxy/peanut shapes. In section \ref{Numerical Technique} we provide the numerical techniques and methods used in this study. We further provide result in section \ref{Results} and discussion is done in section \ref{Discussion}. Finally we summarize and conclude our key results in section \ref{Summary}

\section{Model Setup} \label{Numerical Technique}
\subsection{Initial Galaxy Models}

\begin{table*}
\centering
 \caption{Initial Disk Models with increasing bulge masses}
\label{tab:Model Galaxy}
 \begin{tabular}{lccccccc}
 \hline
 
Models Name & Spin ($\lambda$) & Prograde fraction& Prograde fraction &  Prograde fraction & Prograde fraction & Overall Prograde fraction\\
   & &($R<10$ kpc) &($R<15$ kpc) &($R<20$ kpc) &($R<30$ kpc) & \\
  \hline
  S000 & 0     & 0.36  & 0.41 & 0.43 & 0.45 & 0.48     \\
  S025 & 0.025 & 0.50   & 0.50 & 0.50 & 0.50 & 0.52   \\ 
  S050 & 0.050 & 0.55  & 0.53 & 0.52 & 0.51 & 0.54 \\
  S075 & 0.075 & 0.56  & 0.54 & 0.53 & 0.52 & 0.57 \\
  S100 & 0.1   & 0.73  & 0.65 & 0.61 & 0.59 & 0.62  \\
  SM100 & -0.10  & 0     &  0  & 0.02  & 0.15 & 0.29 \\
  \hline
   \end{tabular}
\begin{flushleft}
Column(1) Model name (2) Spin of the model (3) Fraction of retrograde orbits in dark matter halo within 10 kpc spherical region (4) Fraction of retrograde orbits in dark matter halo within 15 kpc spherical region (5) Fraction of retrograde orbits in dark matter halo within 20 kpc spherical region (6) Fraction of retrograde orbits in dark matter halo within 30 kpc spherical region (7) Fraction of retrograde orbits in all the dark matter halo particles 
\end{flushleft}   

\end{table*}
All the initial galaxy models are generated using GalIC code\citep{Yurin.Springel.2014} which are not aimed particularly at  modeling the Milky Way, although they are similar systems. The code populates orbits of particles in a target stationary potential and achieves the equilibrium N-body realization of the system by solving the collisionless Boltzmann equations. In all of our initial models we have used $10^6$ dark matter halo particles, $10^6$ disk particles.  We also ran the convergence test with the number of particles doubled and found that they show similar results.

\begin{figure*}
    \centering
    \includegraphics[scale=0.85]{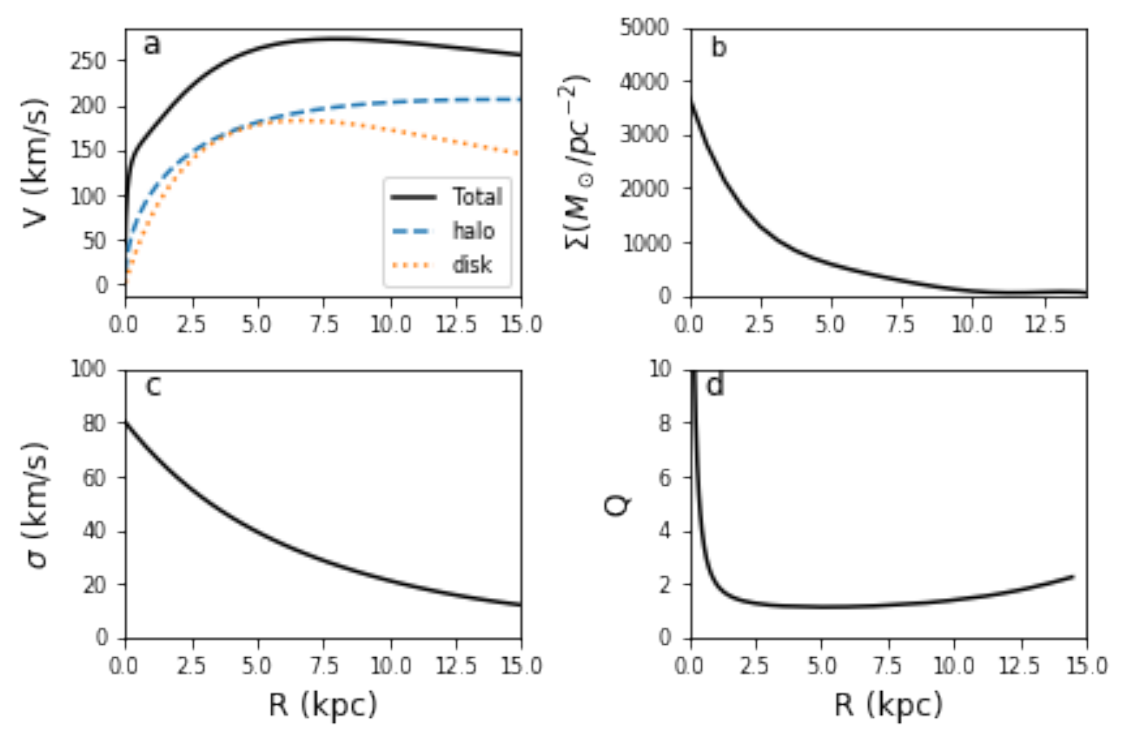}
    \caption{This figure shows radial dependence of the following quantities: a) Circular rotation curve for all the galaxy models along with the contribution from various components b) Surface density of the disk c) velocity of dispersion of stars in the disk d) Toomre parameter of the disk.}
    \label{fig:initial_condition}
\end{figure*}

The total mass ($M_{total}$) of all of our galaxy model is equal to 6.38 x$10^{11} M_{\sun}$. The rotation curve near the sun radius is equal to 250 km/s.  The disks in our galaxy models are locally stable as the value of the Toomre parameter $Q$ is greater than 1. The Toomre factor varies with radius and is given by $Q(R)=\frac{\sigma(R)\kappa(R)}{3.36G\Sigma(R)}$. Here $\sigma(R)$ is the radial dispersion of disk stars, $\kappa(R)$ is the epicyclic frequency of stars and $\Sigma(R)$ is the mass surface density of the disk. We have shown the radial variation of various disk properties in Figure \ref{fig:initial_condition}. 

\begin{figure}
    \centering
    \includegraphics[scale=0.6]{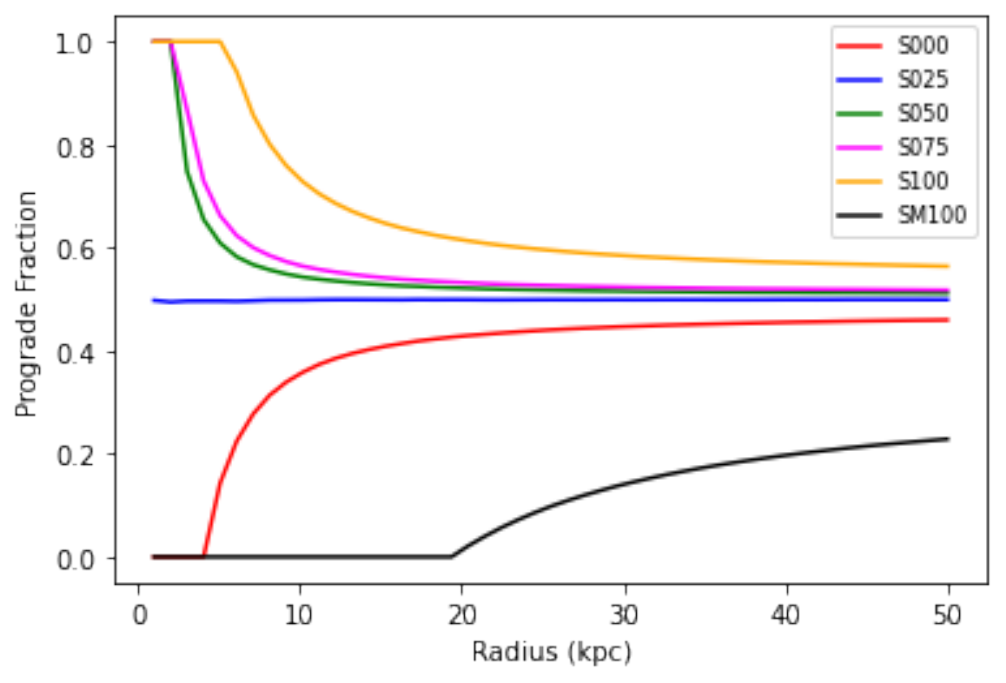}
    \caption{This figure shows the variation of prograde fraction with radius for all our models.}
    \label{fig:prograde_frac}
\end{figure}

The profile of the dark matter halo, which is spherically symmetric, is Hernquist and given as
\begin{equation}
\rho_h=\frac{M_{halo}}{2 \pi} \frac{a}{r(r+a)^3}
\end{equation}   

where $a$ is the scale length of the halo component. This scale length is related to the concentration parameter of the NFW halo by $M_{halo}=M_{200}$ \citep{Springel.etal.2005} so that the inner profile of the halo is identical to the NFW halo. Here $a$ and $c$ are related as follows
\begin{equation}
a=\frac{R_{200}}{c} \sqrt{2[ln(1+c)- c/1+c]}
\end{equation}
where $M_{200}$, $R_{200}$ are the virial mass and virial radius for an NFW halo respectively. $M_{200}$ is equal to 5.68x$10^{11} M_{\sun}$ and $R_{200}$ is equal to 140 kpc. 

The density profile of the disk component has an exponential  distribution in the radial direction and a $\sech^2$ profile in the vertical direction. 

\begin{equation}
\rho_d=\frac{M_d}{4 \pi z_0 h^2} \exp\Bigg(-\frac{R}{R_d}\Bigg) \sech^2\Bigg(\frac{z}{z_0}\Bigg)
\end{equation}
where $R_d=$2.9 kpc and $z_0=$0.58 kpc are the radial scale length and vertical scale length respectively. In our models there is no stellar particle beyond the radius 12$R_D$. Total mass in the disk component is equal to 6.38 x $10^{10} M_{\sun}$= 0.1 $M_{total}$.

For non-spinning halos the total angular momentum of the halo is close to zero where total retrograde fraction of orbits is almost equal to total prograde fraction. For conducting this study we have prepared 5 galaxy models with increasing halo spin where total prograde orbit fractions increases successively. Radial variation of the prograde orbits is shown in Figure \ref{fig:prograde_frac}. In order to increase the spin of the halo we have reversed the direction of retrograde orbits within very central region ($<30$ kpc) which results in increased prograde orbit fractions. We have also an extra model with negative halo spin in which halo orbits are retrograde with respect to disk. The dark matter density remains unvaried under this orbit reversal which is in agreement with Jean's theorem \citep{Jeans.1919J,Binney&Tremaine_2007}. These reversals of orbits maintains the equilibrium distribution of dark matter as the solution of the collisionless boltzmann equation is preserved \citep{Lynden-Bell.1960,Weinberg.1985}. While reversing the direction of halo orbits we have kept the same parameters of the disk. The details of fraction of retrograde orbits in our increasing spin models are given in Table \ref{tab:Model Galaxy}. We can see that halo orbits are generally varied in the central region  and results in different spin models. This is because halo-disk interaction mostly happen in the central region. 

\subsection{Simulation Methods} 

We ran all the N-body simulation using GADGET-2 code \citep{Volker.2005} for evolving our galaxy models. All the galaxy models are evolved up to 9.78 Gyr. The code makes use of the tree method \citep{Barnes.Hut.1986} and computes gravitational forces among all the particles. The time integration of position and velocity of particles is performed using various types of leapfrog methods. The opening angle for the tree is chosen as $\theta_{tot}=$0.4 which has resulted in improved force calculations. The softening length for halo, disk and bulge components have been chosen as 30, 25 and 10 pc respectively. In order to have accurate force calculations, we have used the values of the time step parameter to be $\eta< = $~0.15 and force accuracy parameter $< =$~0.0005 in all of the simulations. As a result in all of our models, the total angular momentum of disk and halo is conserved to within 0.1 $\%$ of the initial value. Throughout the paper we describe our results in terms of code units. Both GalIC and Gadget-2 code have unit mass equal to $10^{10}$  $M_{\sun}$ and unit length is 1 kpc.

\section{Results} \label{Results}
\subsection{Evolution of the Bar Strength} 
There are various ways in which bar strength has been defined in the literature \citep{Combes.Sanders.1981,Athanassoula.2003}. In this study we look for stellar contribution to $m$=2 fourier mode as a proxy for bar strength.

\begin{equation}
a_2(R)=\sum_{i=1}^{N}  m_i \cos(2\phi_i)\\ \hspace{0.5cm}
b_2(R)=\sum_{i=1}^{N} m_i \sin(2 \phi_i)
\label{equation:FM}
\end{equation}

where $a_2$ and $b_2$ are calculated for all the disk particles, $m_i$ is mass of $i^{th}$ star, $\phi_i$ is the azimuthal angle. We have defined the bar strength as  
\begin{equation}
\frac{A_2}{A_0}= \frac{\sqrt{a_2 ^2 +b_2 ^2}}{\sum_{i=1}^{N} m_i} 
\label{eq:barstrength}
\end{equation}
So the bar strength is derived using the cumulative value of $m=2$ Fourier mode for the whole disk which also varies with radius.

\begin{figure}
    \centering
    \includegraphics[scale=0.6]{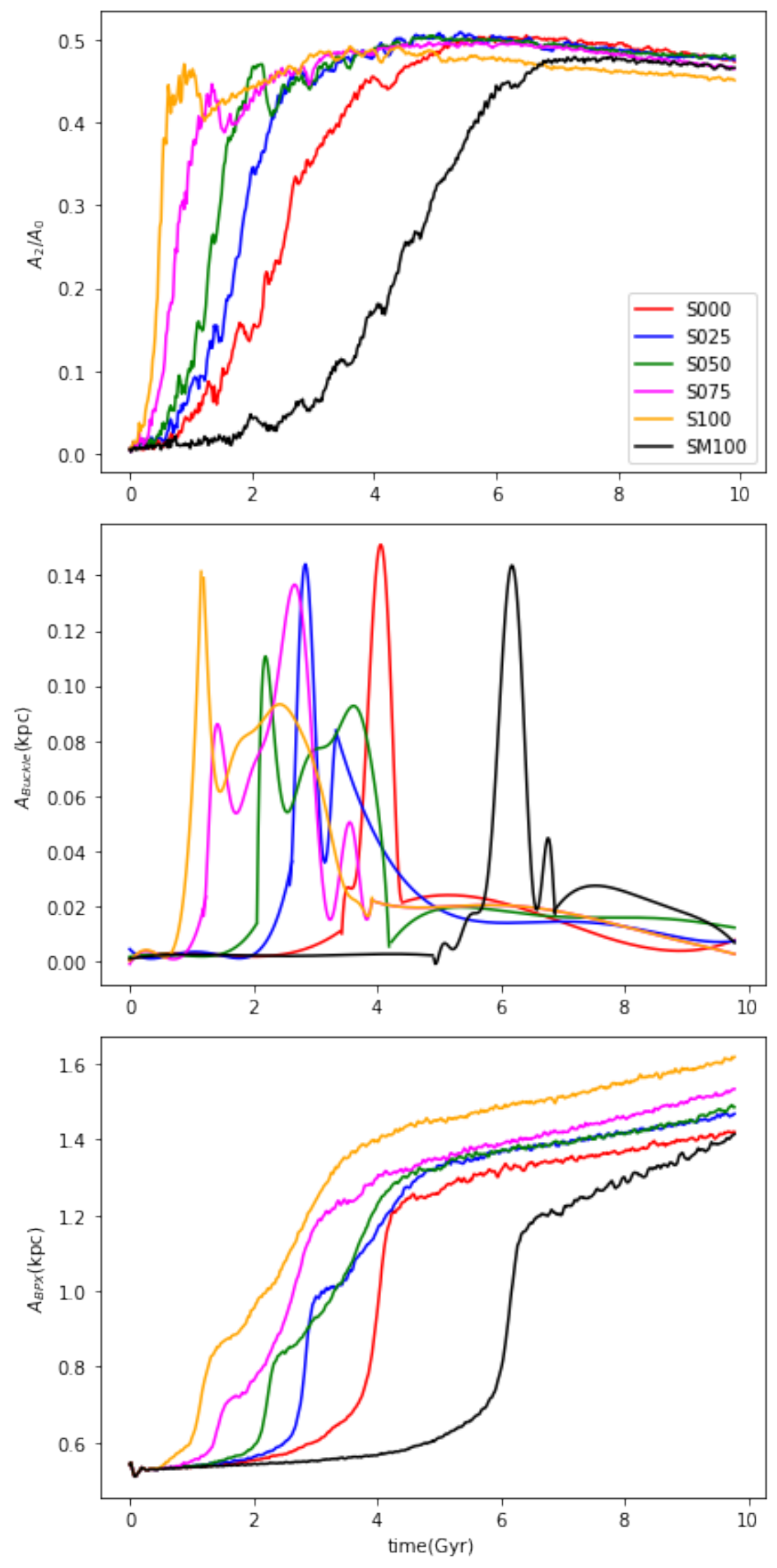}
    \caption{Top panel shows the bar strength evolution with time; middle panel shows the buckling strength evolution with time (here the plots are shifted by arbitrary value for clear comparison) and bottom panel shows evolution of the boxy/peanut strength of bars in all the spinning models.  }
    \label{fig:Bar_BPX_Buckle}
\end{figure}

Bar strength for all of our models has been shown in the top panel of Figure \ref{fig:Bar_BPX_Buckle}. We can clearly see that the bar triggering timescale reduces with the increase in halo spin which is in agreement with earlier studies \citep{Kanak.Saha.Naab.2013,Longetal.2014,Collieretal.2018}. We also find that bar formation in our counter-rotating halo (SM100) has been delayed as shown by earlier studies \citep{Collieretal.2019}. As soon as the bar strength has reached its maximum value it begins to decrease, which is more prominent with increasing halo spin models. This is due to the buckling of bar which we discuss in detail in the next subsection.  Furthermore, as the bar secularly evolves, we find that the bar strength increases with time and finally saturates at the end of the simulation. We find that all the increasing spin models show similar bar strengths at the end of simulations.  This shows that the secular evolution of bars in higher spin models are quite different compared to previously reported studies \citep{Longetal.2014,Collieretal.2018} where the bar becomes weaker with increasing halo spin. We have confirmed our result by using different galaxy initial condition generator; namely Agama  \citep{Eugene.2019}. We explain this difference in  section \ref{Discussion}. 

\subsection{Buckling of bars} 
As all the bars in our simulations buckle as  they evolve. We have measured the time evolution of buckling amplitude which is given by the following equation \citep{Debattista.etal.2006}.

\begin{equation}
    A_{Buckle}= \left | \dfrac{\sum_{i=1}^{N} z_im_i e^{2i\phi_{i}} }{\sum_{i=1}^{N} m_i} \right |
    \label{eq:Buckle}
\end{equation}

where z is the vertical coordinate, $m_i$ is mass of $i^{th}$ star, $\phi_i$ is the azimuthal angle.

The buckling of bar results in the reduced strength of the bars \citep{Raha.etal.1991, MArtinez-Valpuesta.Sholsman.2004}. We find that the reduction in bar strength after buckling increases with the increase in halo spin (top panel of Figure \ref{fig:Bar_BPX_Buckle}). As the bar has buckled in all galaxy models as soon as bar reaches its peak strength, further temporal evolution of bar is defined as the \textit{secular evolution} phase. We have measured the buckling strength using equation \ref{eq:Buckle} which captures the bending of bars very effectively. Time evolution of the buckling strength is shown in the middle panel of Figure \ref{fig:Bar_BPX_Buckle}. We find that as the halo spin increases for all prograde halo models, the buckling of bar happens earlier. While for the counter-rotating halo model (SM100) buckling event is more delayed compared to all prograde spinning halos. These trends are correlated with bar formation timescales as discussed in the previous subsection. We find that the chance of a second buckling event increases with increasing halo spin. The second peak in buckling strength is more pronounced for model S075. 

\begin{figure*}
    \centering
    \includegraphics[scale=1]{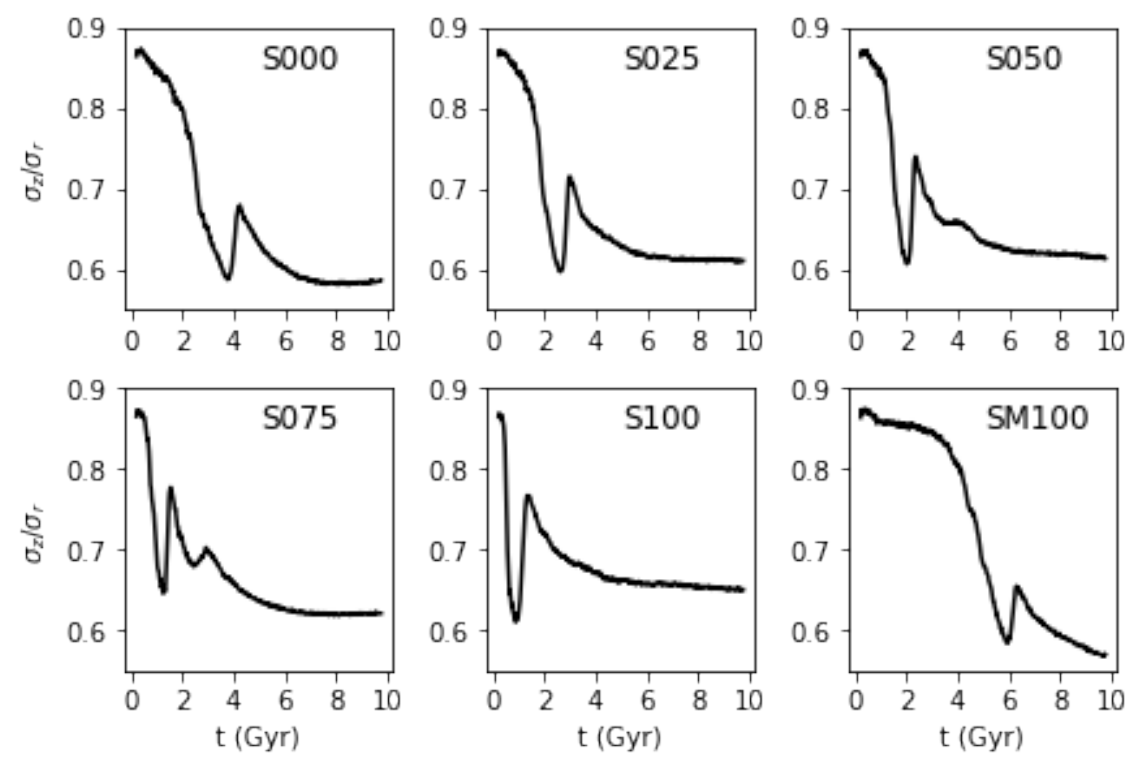}
    \caption{$\sigma_z/\sigma_R $ evolution with time clearly shows that after first jump the tendency of having second jumps increases with increasing spin case. S075 model clearly show second jump indicating second buckling which can be clearly seen in buckling strength (middle panel of fig \ref{fig:Bar_BPX_Buckle}). S100 models doesn't appreciable jump but it shows second buckling.  }
    \label{fig:sigmaz_by_sigmar}
\end{figure*}

 We know from previous studies that buckling also shows signatures in the evolution of $\sigma_z/\sigma_R$ \citep{Lokas.2019}. To capture this we have plotted the time evolution for the same in Figure \ref{fig:sigmaz_by_sigmar}. We can clearly see the sudden jump in  $\sigma_z/\sigma_R$ during the time of buckling for all the models. We notice a few trends clearly; one is that the size in jump increases with increasing halo spin for all prograde orbits. The second trend we notice is that the timescale of these jumps are well correlated with the peak of the buckling strength as seen in the middle panel of Figure \ref{fig:Bar_BPX_Buckle}.  We also notice that the second jump in $\sigma_z/\sigma_R$ tends to be more pronounced with increasing spin, which are seen in S050 and S075 models. We also find that the signature of second buckling in the S075 model  (which shows a second jump) happens around 3 Gyr.

\subsection{Face-on and Edge-on Morphology of bars}
\begin{figure*}
    \centering
    \includegraphics[scale=0.5]{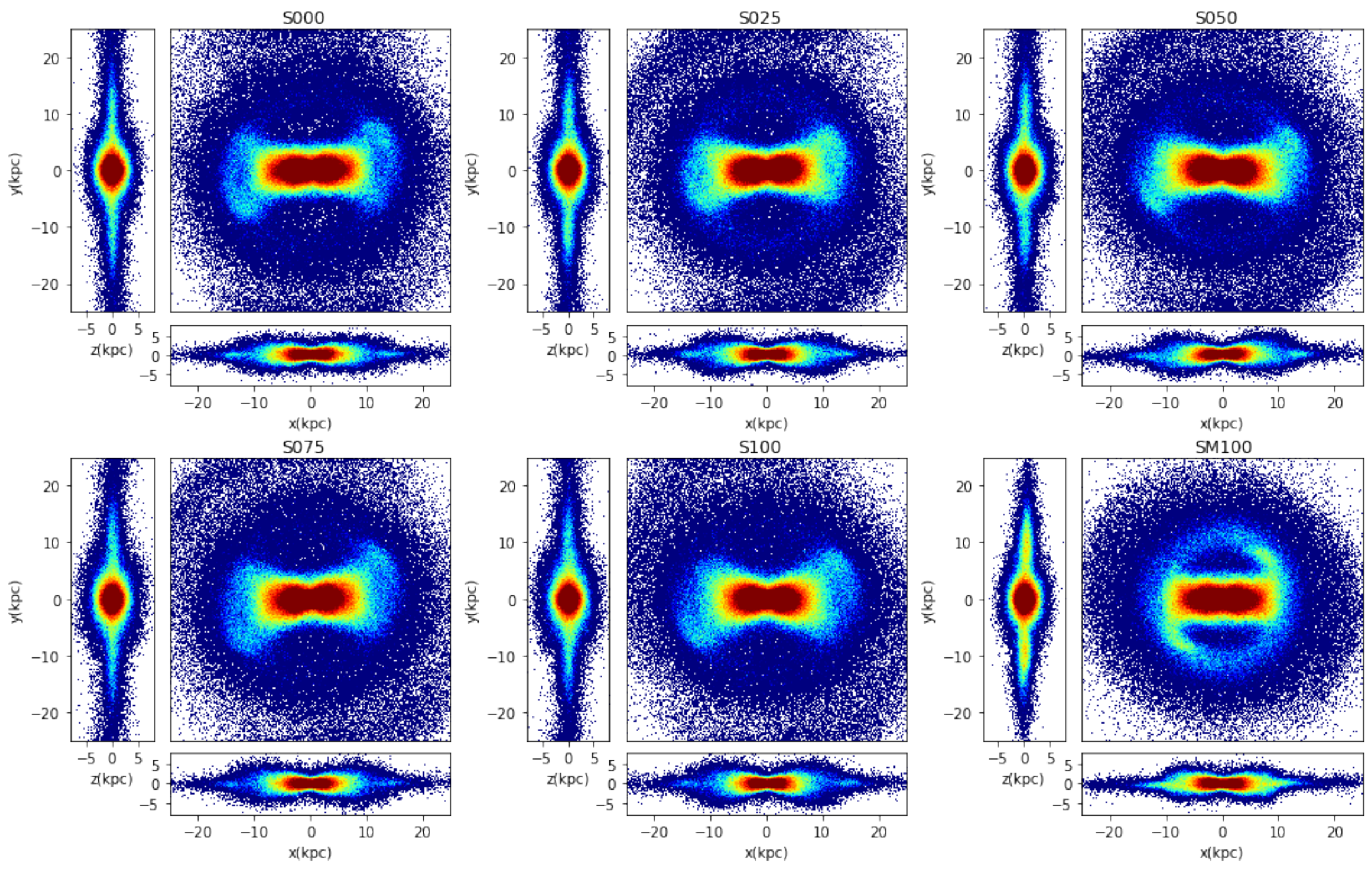}
    \caption{Figure shows density projection maps  at end of simulations (t= 9.78 Gyr) for all the increasing prograde halo spin models ;namely XY, XZ and YZ density maps. XY density map clearly shows that bar changes it's shape from square to farfalle as the halo spin increases. Bottom right maps corresponds to counter-rotating spinning model.}
    \label{fig:projections}
\end{figure*}

As a bar secularly evolves,  its bar strength increases (top panel of Figure \ref{fig:Bar_BPX_Buckle}) which is reflected in decrease in pattern speed ( Figure \ref{fig:Pattern_Speed}). We can clearly see that bar strength for increasing spin models are similar at the end of the simulation. Furthermore, we find that bar morphology changes as the halo spin increases. We have plotted the face-on and edge-on distribution of bars at the end of the simulations in Figure \ref{fig:projections}. We can see clearly that the face-on bar shape changes from rectangular to farfalle with  increasing halo spin for all prograde spin models.  Apart from this we also notice that all of our bars shows ansae (handle) feature which  are quite prominent. These features indicate that all the bars are evolving secularly by capturing orbits from the disk \citep{Martinez-valpuesta.2006}.  

\begin{figure*}
    \centering
    \includegraphics[scale=0.9]{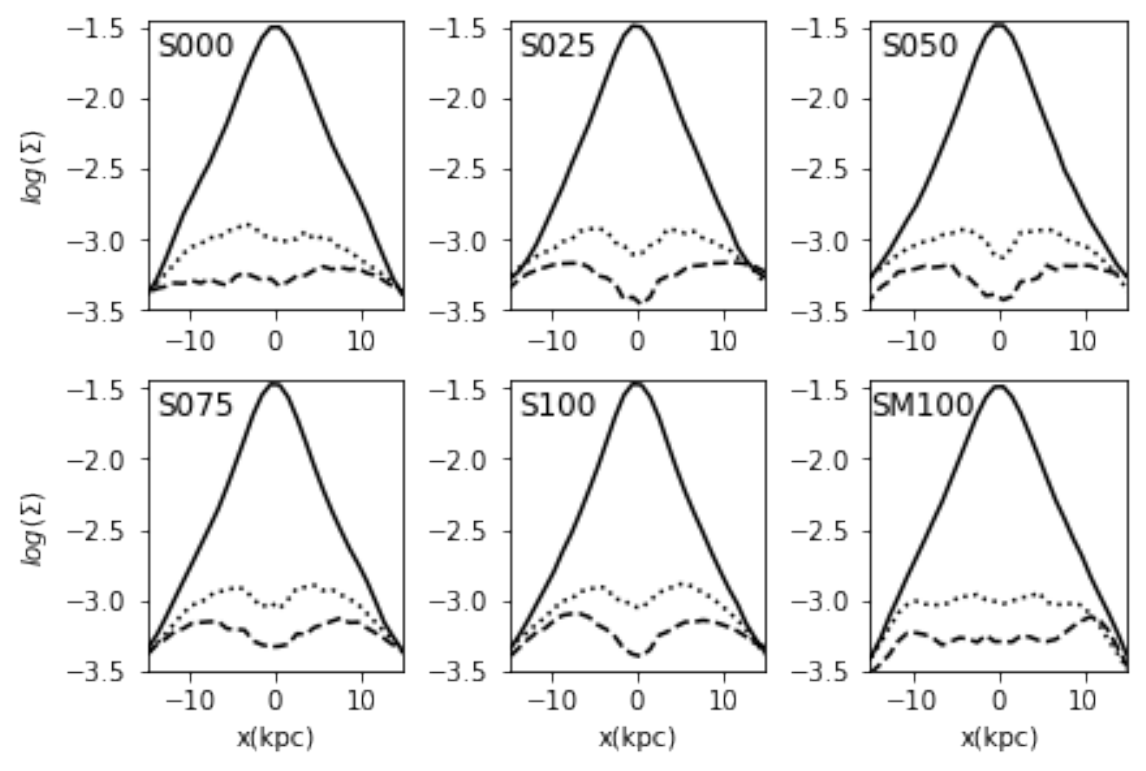}
    \caption{This plots shows the Face-on line profile parallel to the bar major axis at three lines; At the center of bar (Solid line), y=4 kpc line (dashed line) , y=-4 kpc line (dotted line). It shows clearly as the spin of halo increases X-shape feature in Face-on bar becomes more prominent while counter-rotating spin model doesn't show this feature.}
    \label{fig:line_profile}
\end{figure*}

However, for the counter-rotating halo case the bar maintains a box shape
where the spiral is more prominent compared to all prograde spin cases. To quantify the farfalle shape, we further show the line profile along the bar major axis which are passing through the center of bar, i.e. the y= 4 kpc and y=-4 kpc lines in Figure \ref{fig:line_profile}. We can clearly see from these profiles that the bar shows more prominent peanut/X-shapes in higher halo spin models.  However, for the counter-rotating (retrograde) halo spin model (SM100) we see that bar does not show a peanut shape in the face-on profiles.

Figure \ref{fig:projections} also shows the edge-on bars for increasing halo spin models. It is clear from this figure that the peanut/X-shape also gets enhanced as we increase the halo spin. This has also been captured in the $A_{BPX}$ strength which is shown in the bottom panel of Figure \ref{fig:Bar_BPX_Buckle}. The observed shape of the bulge has been suggested to be supported by certain families of orbits mentioned in \citep{Patsis.et.al.2002,Athanassoula.2005}. Both of the box shapes of bars in face-on as well as edge-on maps have been suggested to be supported by quasi-periodic orbits in stability islands or sticky orbits around them \citep{Chaves-Velasquez.et.al.2017}.

\subsection{Boxy/peanut Shape} 

As a result of the buckling a bar grows vertically \citep{Debattista.et.al.2004,Debattista.etal.2006,Martinez-valpuesta.2006,Shen.et.al.2010,Athanassoula.2016, Kataria.Das.2018}. We quantify the boxy/peanut amplitude of the bar which is basically the mean square root of the vertical height of the bar and measured within an annular region of radius ranging from 2 kpc to 8 kpc. We define the buckling strength \citep{Baba.etal.2021}. 
\begin{equation}
    A_{BPX}= \sqrt{\dfrac{\sum_{k=1}^{N} m_k z_k^2 }{\sum_{k=1}^{N} m_k}} 
    \label{eq:BPX}
\end{equation}

where $z$ is the vertical coordinate, $m_k$ is mass of $k^{th}$ star.

As the bar evolves secularly it  becomes vertically thicker due to mainly three ways as discussed in \citet{Sellwood.Gerhard.2020}, which are buckling instability, heating of bar orbits while a passage through 2:1 vertical resonances and a gradual capturing of bar orbits by 2:1 resonances. We have captured the thickening of bars using boxy/peanut shape strength with time in the lower of the Figure \ref{fig:Bar_BPX_Buckle}. We see that evolution of $A_{BPX}$ with time shows varied nature as the halo spin increases. In all of the spinning halo models, there is a steep increase in $A_{BPX}$ at the time of buckling which decreases with increase in halo spin. After this steep increase there is a gradual increase in $A_{BPX}$, slope of this gradual increase, increases with spin of the halo which is quite clear as we move from S000 to S075 models. We find that the steepest gradual slope for Model S075 corroborates with a second buckling event. Furthermore, we observe that $A_{BPX}$ increases more gradually for all the prograde spinning models and the final $A_{BPX}$ shows a monotonic increase in $A_{BPX}$ with halo spin. $A_{BPX}$ is around 20$\%$ higher in the highest spin model (S075) compared to the lowest one (S000). For the counter-rotating spin model (SM100) we also find that $A_{BPX}$ increases steeply during buckling and then it increases gradually.  It has a much higher slope compared to the slope of gradual $A_{BPX}$ increase in all the prograde spin models.

\subsection{Bar Pattern Speed and R Parameter} \label{Pattern_R}
Pattern speed ($\Omega_P$) of the bar has been measured as the change in phase angle  $\phi=\frac{1}{2}\tan^{-1}\bigg(\frac{b_2}{a_2}\bigg)$ of the bar with time where $a_2$ and $b_2$ are defined by equation \ref{equation:FM}. This is measured using the Fourier modes ($m=2$) in concentric annular bins of size 1 kpc throughout the disk of the galaxy. In order to measure the phase angle of the bar we have used the Fourier mode of the  annular region corresponding to the peak in maximum value of $A_2/A_0$. 

Time evolution of bar pattern speed has been shown in Figure \ref{fig:Pattern_Speed}. We can clearly see that the initial bar pattern speed increases with increase in halo spin. As the bar evolves secularly, the bar pattern speed decreases due to the dynamical friction of the halo \citep{Debattista.Sellwood.2000}. We notice that the final bar pattern speed for all the increasing spin models saturate to similar values  and are correlated with the saturation in bar strengths (top panel of Figure \ref{fig:Bar_BPX_Buckle}) at the end of the simulations. The bar pattern speed for higher spin halos reduces sharply in the beginning compared to lower spin halos, although the trend reverses at the end of the simulations. The retrograde halo spin model (SM100) shows the fastest decrease in pattern speed compared to all prograde halo spin models.  This clearly indicates  that the dynamical friction by the counter-rotating halo is stronger compared to the prograde one.   

 The R parameter is defined as the ratio of corotation radius ($R_{CR}$)  to bar major axis ($R_b$). Corotation radius is  measured as the distance of the Lagrange points (L1/L2) from the center of disk. It is well known that in the bar rotating frame the centripetal force is balanced by the gravitational force at the Lagrange points, therefore the effective force is zero \citep{Debattista.Sellwood.2000}. We evaluate the gravitational forces and centripetal forces along the bar major axis within the radial bin of 1 kpc size. We measure the L1/L2 Lagrange point as the radius of bin where both gravitational force and centripetal forces are equal.

Bar lengths have been measured using several methods in the literature. Some of these use the ellipicity and PA variation along the fitted isophotes \citep{Erwin_2003,Erwin.2005,Marinova_Jogee_2007,Zou.Shen.2014,Kataria.Das.2019} while other studies \citep{Rosa-Guevara.et.al.2020} use the radial variation of bar strength (as defined in equation \ref{eq:barstrength}) and phase angle of the bar ($\phi$). In this study we have used average of bar lengths from three techniques namely $\dfrac{A_2}{A_0}$ peak methods, $a_5$ and $a_{min}$ (for more details see Appendix \ref{Barlength}) to measure the length of the bar. We also show the visual overlay of bar lengths measured from various methods on the density contours of bar (S000 model) in Appendix \ref{Barlength}. 

 Table \ref{table:R} shows the values of bar lengths from various measurement methods, corotation radius and R parameter for all models. We see that all the bars in increasing halo spin models have roughly similar sizes, while the counter-rotating halo models (SM100) have slightly smaller bars. We  do not find any correlation between the corotation radius and increasing halo spin.  However, we do see that all prograde halos as well the retrograde halo show slow bars (R $>>$1.4). These slow bars are a result of the dynamical friction due to the surrounding dark matter halos, which is also similar irrespective of halo spin. We also find the slowest bar is in the case of the counter-rotating halo where the bar has reduced its pattern speed despite  being triggered late (Fig \ref{fig:Pattern_Speed}).

\begin{table*}
\centering
 \caption{R parameter of all the barred models}
 \begin{tabular}{lccccccc}
 \hline
 
Models Name & Bar length  & Bar length  & Bar length  & Avg. Bar length & Corotation Radius & R parameter \big($\dfrac{R_{CR}}{R_b}$\big) \\
    &($\dfrac{A_2}{A_0}$ peak)& ($a_5$) &($a_{min}$) & ($R_b$) &($R_{CR}$)& \\

    &(kpc)& (kpc) &(kpc) & (kpc)&(kpc)& \\ 
   
  \hline
  S000 & 7.25 & 12.40 & 17.43& 12.55 & 26.75  & 2.13   \\
  S025 & 5.5 & 10.89 & 16.42 & 10.94 & 25.75   & 2.35  \\ 
  S050 & 6 & 10.38 & 17.09 & 11.16  & 25.75  & 2.31   \\
  S075 & 7.25& 11.14& 18.04 &12.14 & 28.25  & 2.33   \\
  S100 & 6.25 &10.89& 18.04 & 11.73 & 27.75  & 2.37     \\
  SM100& 6.25& 8.86 & 12.45& 9.19  & 24.25 & 2.63    \\
  \hline
  \label{table:R}
   \end{tabular}
\begin{flushleft}
Column(1) Model name (2) Bar Radius measured using $\dfrac{A_2}{A_0}$ peak method (3) Bar Radius measured using bar phase angle variation by 5 degrees ($a_5$) (4) Bar radius measured using minimum ellipticity of isophotes of increasing radii ($a_{min}$). (5) Average bar radius of $\dfrac{A_2}{A_0}$, $a_5$ and $a_{min}$ (6) Corotation radius measured at t=9.78 Gyr. (7) R parameter which is ratio of corotation radius to bar radius. 
\end{flushleft}   

\end{table*}

\subsection{Angular Momentum Transport}

\begin{figure}
    \centering
    \includegraphics[scale=0.6]{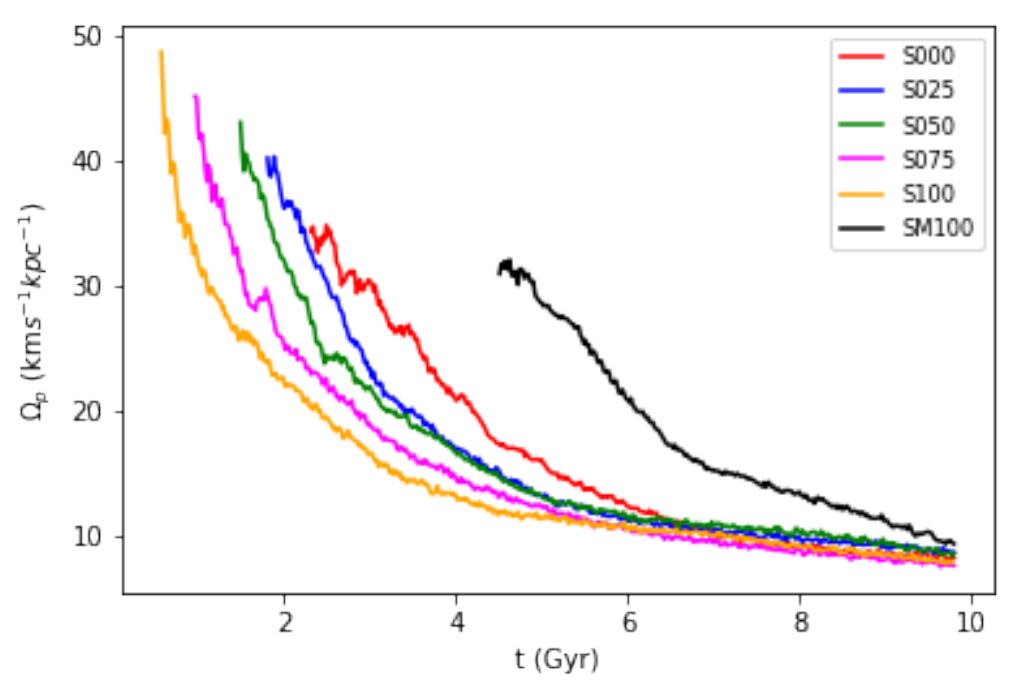}
    \caption{Evolution of the bar pattern speed for all the spinning halo models are shown hereby. }
    \label{fig:Pattern_Speed}
\end{figure}

\begin{figure}
    \centering
    \includegraphics[scale=0.6]{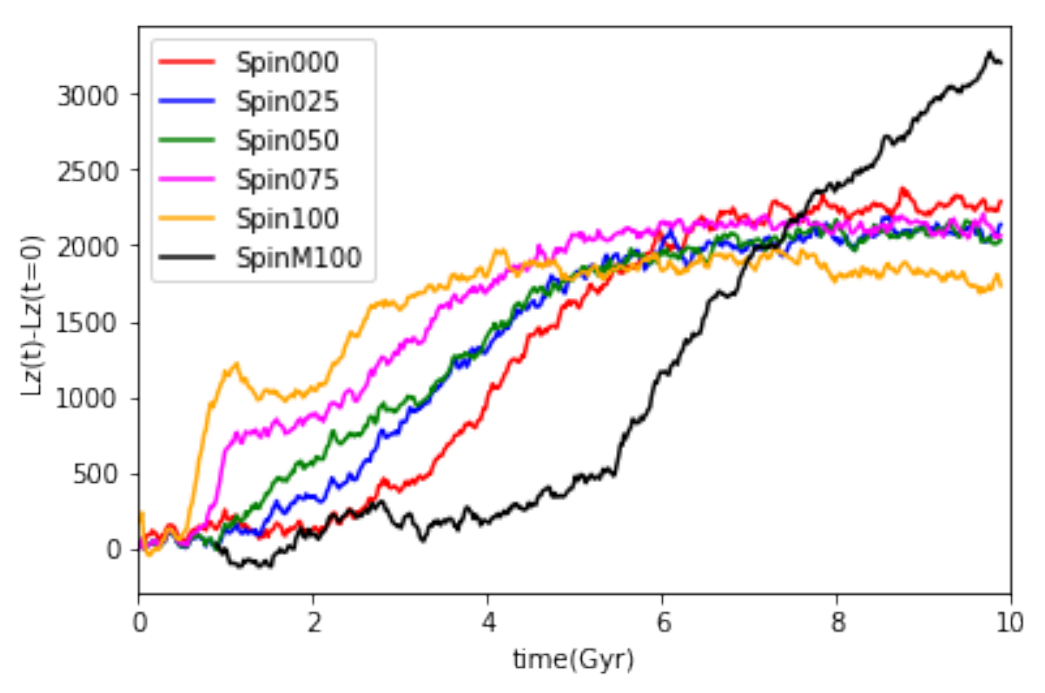}
    \caption{Angular momentum gained by halos within 20 kpc cylindrical region for all the spinning halo models. Faster spinning halos larger angular momentum in the triggering phase of bar while trend reverses during secular evolution.}
    \label{fig:AM}
\end{figure}

As a bar evolves it transfers angular momentum from the disk to the surrounding dark matter halo. We have plotted the evolution of the increase in the angular momentum of the halo within a 20 kpc cylindrical region in Figure \ref{fig:AM}. We can clearly see that the higher spin halo gains more angular momentum earlier compared to the lower spin halos during the bar triggering phase. In this bar triggering phase, the rate of angular momentum change is also higher for high spin halos. As the bars secularly evolve, we find that the rate of angular momentum transfer to the halo increases for low halo spin cases.  The rate of angular momentum transfer in the bar triggering phase as well as in the secular evolution phase explains the pattern speed behaviour respectively (Figure \ref{fig:Pattern_Speed}). We can clearly see that the retrograde spin model (SM100) gains angular momentum with the highest rate and the final gain at the end of the simulation is much larger than all the prograde halo spin models. Furthermore, the highest gain in angular momentum is clearly captured in the faster decrease in pattern speed for the counter-rotating case.

\section{Discussion} \label{Discussion}

We see that the prograde halos with increasing spin trigger bars earlier as  also seen in previous studies \citep{Kanak.Saha.Naab.2013, Longetal.2014}. While in the secular evolution we find that bars keep on growing irrespective of halo spin which is different from earlier  studies \citep{Longetal.2014,Collieretal.2018}. We attribute this change to difference in the nature of angular momentum distribution for a given spin of the halo. As we know, the halo spin parameter depends on the total angular momentum of the surrounding dark matter halo as shown in equation \ref{eq:lambda}. Since the halo is distributed over a large radial scale compared to the disk, the total angular momentum of the halo can  correspond to different radial distributions of internal halo angular momenta for a given  total spin. In this  study we increased the angular momentum in the central region of halos to obtain fast spinning halos as shown in Table \ref{tab:Model Galaxy} which is represented in terms of decreasing retrograde orbits in the central region. This is because most of the disk-halo angular momentum interaction happen within the bar region. \cite{Collieretal.2018} has shown that in high halo spin models, bar regrowth after buckling is inhibited by the collusion of two factors  namely (i)~bars becoming weaker and (ii)~ the inability of halo to absorb angular momentum. In this study we clearly see that the buckling of a bar does not weaken the bar significantly (top panel of Figure \ref{fig:Bar_BPX_Buckle}), which is the case in previous studies \citep{Longetal.2014,Collieretal.2018}.  Therefore strong bars are still torqued down due to dynamical friction of the halos despite the higher halo spin. Therefore bars can still exchange angular momentum and grow with increasing halo spin, although the rate of transfer reduces with increasing spin (Figure \ref{fig:AM}) which is in agreement with previous studies. We further conduct several tests using initial conditions generated from other initial condition generators i.e Agama \citep{Eugene.2019}. We confirm our results with new models that are shown in Appendix \ref{Agama_IC}.

\begin{figure}
    \centering
    \includegraphics[scale=0.65]{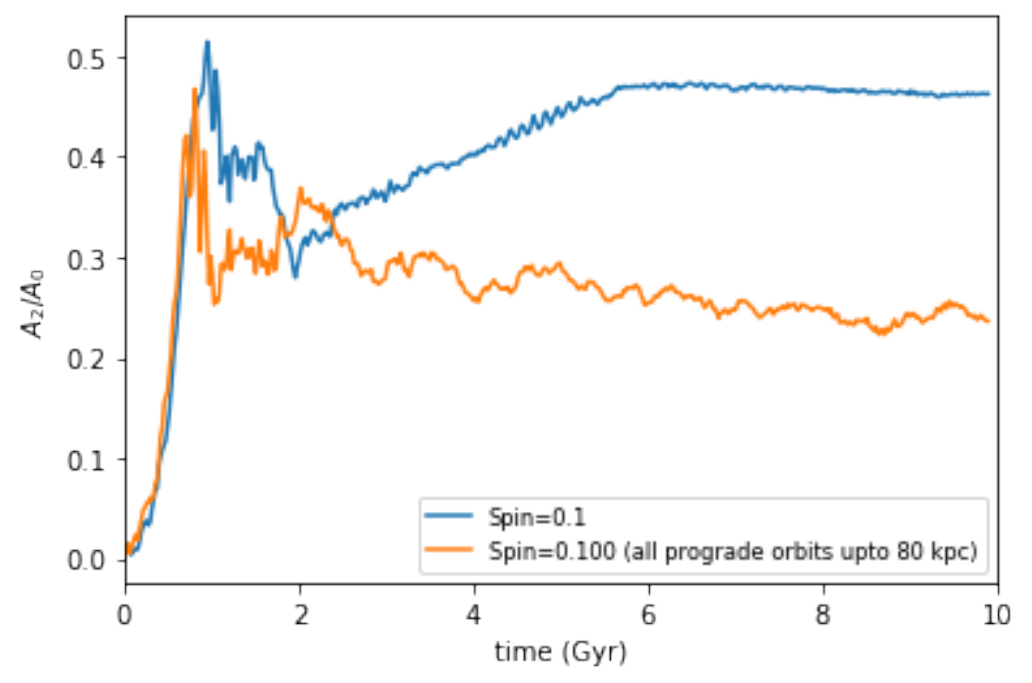}
    \caption{This figure shows time evolution of bar strength for the high spin models (Spin100). This clearly shows dark matter halo having all prograde orbits within 80 kpc radius region suppresses the bar strength in secular evolution phase as seen in previous studies\citep{Longetal.2014,Collieretal.2018}.}
    \label{fig:Agama_high_spin_Exp}
\end{figure}

We further test our high spin model (Spin100) by increasing the prograde orbits of dark matter halo to larger radii. We find that Spin100 galaxy model with all the prograde orbits or no retrograde orbits up to 80 kpc doesn't show bar growth in secular evolution compared to the model with retrograde fraction distribution as shown in table \ref{tab:Model Galaxy}. Figure \ref{fig:Agama_high_spin_Exp} clearly show that bar formation timescale and peak bar strength are similar for both the models. In the secular evolution bar formation is suppressed in model with all the prograde orbits of halo within 80 kpc radii as seen in previous studies \citep{Longetal.2014,Collieretal.2018}. This is because the bar is no longer able to transport angular momentum to halo in secular evolution which is not the case with the model with lower prograde fraction in central region. Hence, the importance of inner halo angular momentum ( $<$ 30 kpc) is highlighted by this study which doesn't allow bar weakening in secular evolution, even for the high spin cases. Given the richness in radial variation of the angular momentum of halo for a given halo spin, we show the importance of initial condition in another study (Kataria et al. in preparation).

 From a theoretical point of view it is still not clear what causes strong buckling and the recurrence of buckling events. It has been shown that buckling events in bars are suppressed due to the addition of nuclear mass in the disk \cite{Sellwood.Gerhard.2020} and  a larger gas component \citep{Berentzen.et.al.1998,Debattista.etal.2006,Woznaik&Michel-Dansac.2009,Lokas.2020}. In this study we find that buckling becomes more prominent with the increasing spin of prograde halos, which causes a large decrease in bar strength (top panel of Figure \ref{fig:Bar_BPX_Buckle}) and an increase in the sudden jump of $\sigma_z/\sigma_R$ (Figure \ref{fig:sigmaz_by_sigmar}). \cite{Martinez-valpuesta.2006} has shown the presence of a recurrent buckling event given that the bar strength keeps on increasing after the first buckling. We also find that bar strength increases after the first buckling and  that the chance of having a second buckling becomes larger with increasing spin of the prograde halo.      

\cite{Sellwood.Gerhard.2020} discusses the three mechanisms of bar thickening which are quite relevant to our work. In our models we find that various phases of bar thickening as seen in boxy/peanut strength $A_{BPX}$ (bottom panel of \ref{fig:Bar_BPX_Buckle}). In our prograde high halo spin models we find that there are three phases of bar thickening for which the slope of $A_{BPX}$ decreases in successive phase. First phase corresponds to sharp increase in bar thickening as a result of buckling. Second phase of bar thickening points towards the second buckling event which is less steeper than first phase. Finally the third phase where bar thickens at a much slower pace points towards the mechanism of capture or passage of bar orbits by 2:1 vertical resonance.

\section{Summary and Conclusion} \label{Summary}

In this article we have conducted N-body experiments to look into the effect of increasing halo spin on bar secular evolution which results  in the formation of boxy/peanut shaped bars during their evolution. These increasing spin models corresponds to increasing angular momentum in the very central regions ($<$30 kpc) of the galaxy. We also look at a retrograde spinning halo model which is counter-rotating with respect to the disk. We confirm that bars are triggered earlier in the faster spinning prograde halos while it is delayed in the counter-rotating halos as shown in previous studies \citep{Collieretal.2018,Collieretal.2019}. We summarize our results below.
\begin{itemize}
    \item We find that secularly evolved bars in high spinning prograde halos remain strong until 9.78 Gyr of evolution. We also report that the mild buckling event soon after bar strength peaks where bar strength does not change for less than 2$\%$ during buckling.
    \item We find that the face-on shape of the bar changes from rectangular to pronounced farfalle shape as the spin of halo increases at the end of secular evolution phase.
    \item The boxy/peanut shapes of bars gets more pronounced as halo spin increases. Furthermore, we quantify that there is around $20 \%$ increase in boxy/peanut strength ($A_{BPX}$ for the lowest (S000) to highest (S100) spin models in our study.
    \item We find that evolving bars with increasing halo spins buckle earlier and  have a tendency to undergo a second buckling event, which is  enhanced with increasing halo spin. The second buckling events are captured by buckling strength ($A_{Buckle}) $ and $\sigma_z/\sigma_R$ ratio. We also find that the timescale of the second buckling event is larger (order of a Gyr) than the first buckling event (~150 Myr).
       
    \item  We conclude that all the bars in our simulations are slow for which R $\gg$ 1.4. The R parameters for all the simulations do not depend much on halo spin in our simulations. 
    
\end{itemize}

\section*{Acknowledgements}

The research presented here is partially supported by the National Key R\&D Program of China under grant No. 2018YFA0404501; by the National Natural Science Foundation of China under grant Nos. 12025302, 11773052, 11761131016; by the ``111'' Project of the Ministry of Education of China under grant No. B20019; and by the China Manned Space Project under grant No. CMS-CSST-2021-B03. J.S. also acknowledges support from a \textit{Newton Advanced Fellowship} awarded by the Royal Society and the Newton Fund. We thank Volker Springel for Gadget code which we used to run our simulations. We thank Jerry Sellwood, Mousumi Das, Victor Debattista, Zhi Li for scientific discussion on initial condition of galaxies which helped to progress in this work.  This work made use of the Gravity Supercomputer at the Department of Astronomy, Shanghai Jiao Tong University, and the facilities of the Center for High Performance Computing at Shanghai Astronomical Observatory. We have also used 'NOVA" High Performance Computing facility at Indian Institute of Astrophysics. 
software used: numpy \citep{Harris.et.al.2020}, matplotlib \citep{Hunter.2007}, pynbody \citep{Pynbody.2013} and astropy \citep{Astrop.collaboration.2018} packages. 
\section*{Data Availability}

All the simulated data which is generated during this work will be provided on reasonable request to the corresponding author.


\bibliography{sample631}{}
\bibliographystyle{aasjournal}


\appendix

\section{Measurements of the bar length} \label{Barlength}

\begin{figure*}
    \centering
    \includegraphics[scale=0.35]{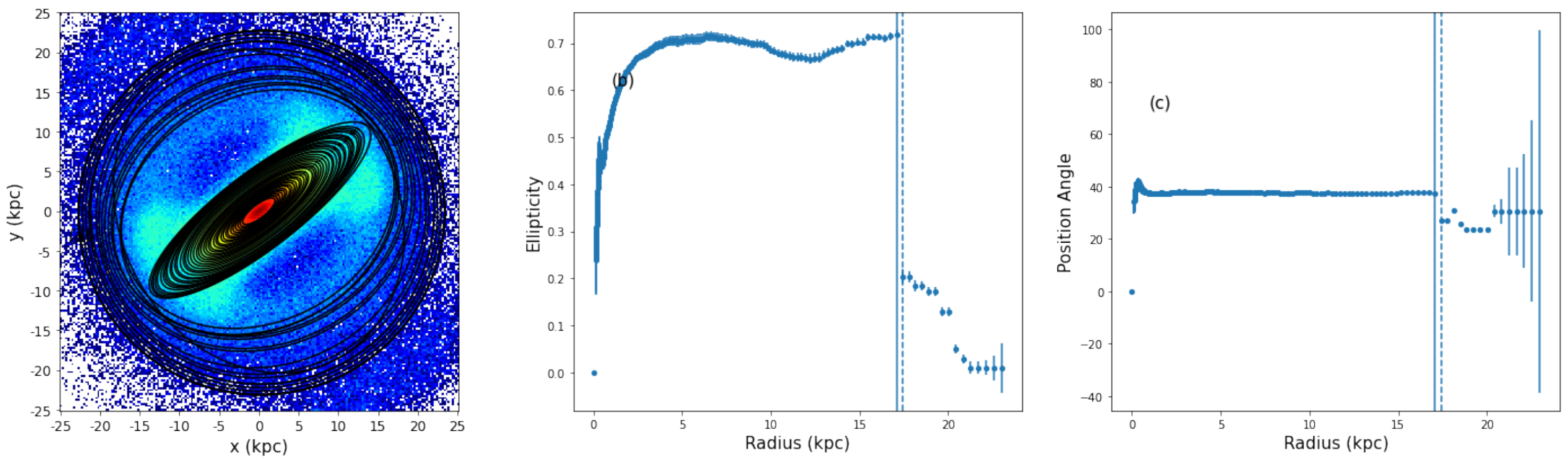}
    \caption{Left panel shows the fitted isophotes with increasing semi-major axis along with density map of face-on disk; middle panel shows the ellipticity of all the isophotes with increasing semi-major axis; right panel shows the position angle of the isophotes with increasing semi-major axis. Here the solid lines correspond to $a_{max}$ and dashed lines corresponds to $a_{min}$ which is denoting the length of bars with maximum ellipticity and minimum ellipticity values such that postion angle of isophotes change within 10 degrees. }
    \label{fig:isophotel}
\end{figure*}

In this appendix we show the comparison of various methods used to measure bar length as discussed in subsection \ref{Pattern_R}. We show only S000 model though we have conducted the similar analysis for all the models.  Figure \ref{fig:isophotel} shows fitted isophotes with increasing semimajor axis using $photutils$ package \citep{Astropy.2018}. Then we calculate the bar lengths using various definitions like $a_{min}$ and $a_{max}$ which corresponds to maximum and minimum ellipticity as the position angle isophotes change within 10 degrees. 

\begin{figure}
    \centering
    \includegraphics[scale=0.55]{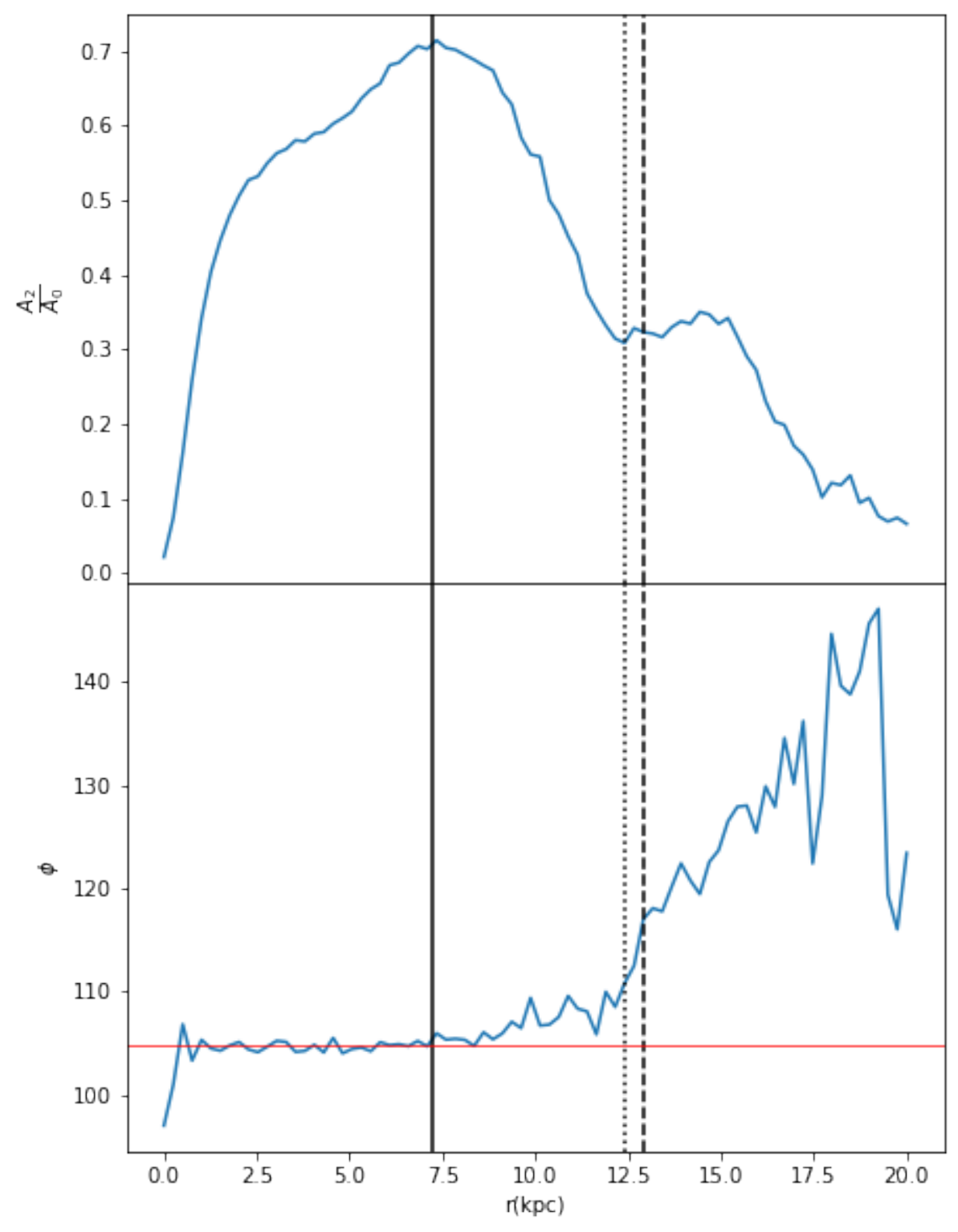}
    \caption{Top panel shows the variation of $\dfrac{A_2}{A_0}$ with radius;  bottom panel shows the variation of the bar phase angle ($\phi$) with radius. The solid line corresponds to radius at which $\dfrac{A_2}{A_0}$ peaks, dotted and dashed line corresponds to radii at which bar phase angle ($\phi$) deviates from constant value by 5 degrees and 10 degrees respectively. }
    \label{fig:BS radial variation}
\end{figure}

\begin{figure}
    \centering
    \includegraphics[scale=0.65]{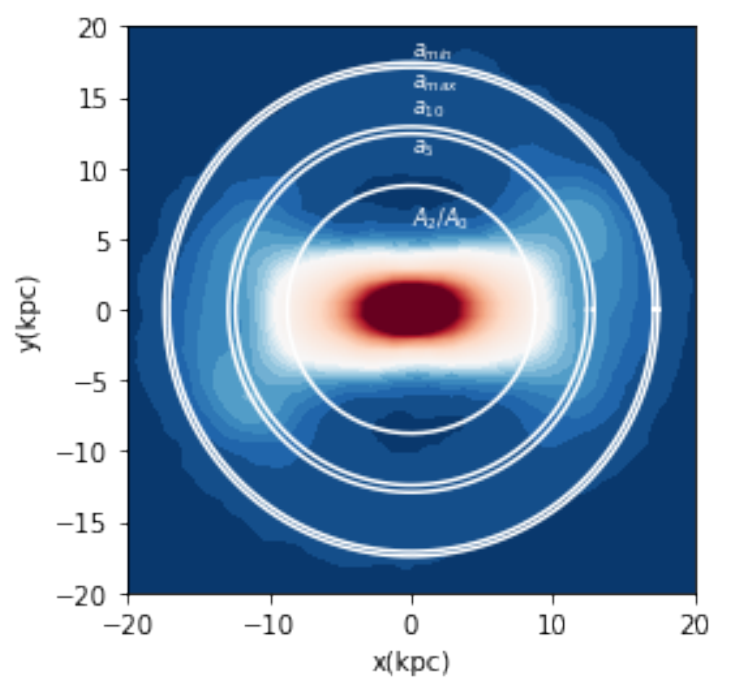}
    \caption{This figure shows the visual overlay of circles, with radii representing bar lengths measured using various methods, on the contour plot of bar in S000 model. The methods includes $\dfrac{A_2}{A_0}$ peak technique, $a_5$ and $a_{10}$ as mentioned in Fig. \ref{fig:BS radial variation}, $a_{min}$ and $a_{max}$ mentioned in Fig. \ref{fig:isophotel}}
    \label{fig:Barlength_vis}
\end{figure}

We further measure the bar strength from radial variation of bar amplitude. We have looked for $\dfrac{A_2}{A_0}$  and bar phase angle ($\phi$) variation with radius as shown in Figure \ref{fig:BS radial variation}. Here $\dfrac{A_2}{A_0}$ is measured in radial bins of 1 kpc size using equation \ref{eq:barstrength}. We measure the bar length as the radius where  $\dfrac{A_2}{A_0}$ peak. Bar phase angle ($\phi$) is measured using fourier components of bar strength in radial bins of 1 kpc size. We measure bar length namely $a_5$ and $a_{10}$ as the radii where $\phi$ deviates from a constant value by 5 degrees and 10 degrees.
When compared visually (Figure \ref{fig:Barlength_vis}) we find that both  $a_{min}$ and $a_{max}$ overestimate the bar length while $\dfrac{A_2}{A_0}$ peak method under estimate the bar length. 
Bar lengths $a_5$ and $a_{10}$ measured using variation of bar phase angle provide visually reasonable estimates.
In this study we use barlength which is average of three methods namely $\dfrac{A_2}{A_0}$ peak, $a_{5}$ and $a_{min}$.

\section{Testing our results with other Galaxy initial Condition generator} \label{Agama_IC}
\begin{figure}
    \centering
    \includegraphics[scale=0.65]{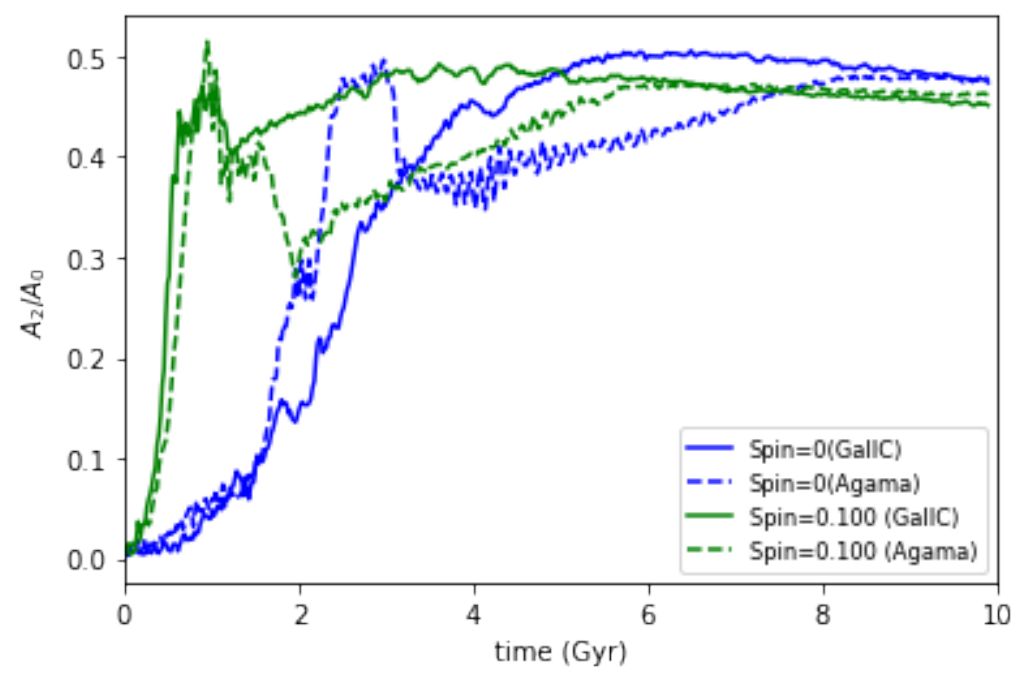}
    \caption{This figure shows the bar strength time evolution for lowest and highest spin models. Solid line models are generated using GalIC initial condition generator while dashed model are generated using Agama initial condition generators.}
    \label{fig:Agama_GalIC_Comparison}
\end{figure}

We test the robustness of our result by using galaxy models generated from $Agama$ initial condition generator \citep{Eugene.2019}. $Agama$ realizes N-body equilibrium system in an action-based approach. Our $Agama$ models are having same density distribution for disk and halo components while kinematics are closer to GalIC models. We also make sure that $Agama$ models have same fractions of retrograde orbits for given spin of dark matter halo component as shown in table \ref{tab:Model Galaxy}. Figure \ref{fig:Agama_GalIC_Comparison} shows the comparison of bar strength evolution with time for lowest (Spin000) and highest (Spin100) models generated using GalIC and Agama initial condition generators. We can see the bar formation timescales are similar for a given spin though the bar strength evolution is not exactly same given slight difference in the kinematics. We confirm that bar forms earlier with increasing halo spin with Agama models also. Further same models also show secular growth for high spin models and the bar is not suppressed in the secular evolution phase.

\end{document}